\pdfoutput=1
\documentclass{pnastwo}

\usepackage{graphicx}

\usepackage{amssymb,amsfonts,amsmath}

\DeclareMathOperator{\erfc}{erfc}
\DeclareMathOperator{\erf}{erf}
\DeclareMathOperator{\Tr}{Tr}

\begin{document}

%----------------------------------------------------------------------------------------
%	TITLE AND AUTHORS
%----------------------------------------------------------------------------------------

\title{Nonlinear Analogue of the May-Wigner Instability Transition} % For titles, only capitalize the first letter

%------------------------------------------------

\author{Yan V. Fyodorov\affil{1}{Queen Mary University of London, School of Mathematical Sciences, London E1 4NS, United Kingdom}, \and
Boris A. Khoruzhenko\affil{1}{Queen Mary University of London, School of Mathematical Sciences, London E1 4NS, United Kingdom}
}
\contributor{Submitted to Proceedings of the National Academy of Sciences of the United States of America}

%%%Newly updated.
%%% If significance statement need, then can use the below command otherwise just delete it.
%\significancetext
%\significancetext{Complex systems equipped with stability feedback mechanisms become unstable to small displacements from equilibria as the complexity (as measured by the interaction strength and number of degrees of freedom) increases. This paper takes a global view on this transition from stability to instability. Our examination of a generic dynamical system whereby $N$ degrees of freedom are coupled randomly shows that the phase portrait of complex multi-component systems undergoes a sharp transition as the complexity increases. The transition manifests itself in the exponential explosion of the number of equilibria and the transition threshold is quantitatively similar to the local instability threshold. Our model provides a mathematical framework for studying generic features of the global dynamics of large complex systems.}

%\maketitle % The \maketitle command is necessary to build the title page

%----------------------------------------------------------------------------------------

\maketitle % The \maketitle command is necessary to build the title page

\begin{article}

%----------------------------------------------------------------------------------------
%	ABSTRACT, KEYWORDS AND ABBREVIATIONS
%----------------------------------------------------------------------------------------

\begin{abstract}
{We study a system of $N\gg 1$ degrees of freedom coupled via a smooth homogeneous Gaussian vector field with both gradient and divergence-free components. In the absence of coupling, the system is exponentially relaxing to an equilibrium with rate $\mu$. We show that, while increasing the ratio of the coupling strength to the relaxation rate, the system experiences an abrupt transition from a topologically trivial phase portrait with a single equilibrium into a topologically non-trivial regime characterised by an exponential number of equilibria, the vast majority of which are expected to be unstable. It is suggested that this picture provides a global view on the nature of the May-Wigner instability transition originally discovered by local linear stability analysis.
}
\end{abstract}

%------------------------------------------------


\dropcap{W}ill diversity make a food chain more or less stable? The prevailing view in the mid-twentieth century was that diverse ecosystems have greater resilience to recover from events displacing the system from equilibrium and hence are more stable. This `ecological intuition' was challenged by Robert May in 1972 \cite{May1972}. At that time, computer simulations suggested that large complex systems assembled at random might become unstable as the system complexity increases \cite{GA1970}. May's 1972 paper complemented that work with an analytic investigation of the neighbourhood stability of a model ecosystem whereby $N$ species at equilibrium are subject to random interactions.

The time evolution of large complex systems, of which model ecosystems is one example, is often described within the general mathematical framework of coupled first-order nonlinear ordinary differential equations (ODEs). In the context of generic systems, the Hartman-Grobner theorem then asserts that the neighbourhood  stability of a typical equilibrium can be studied by replacing the non-linear interaction functions near the equilibrium with their linear approximations. It is along these lines that May suggested to look at the linear model
\begin{equation}\label{May1}
\frac{d y_j}{dt}=-\mu y_j +\sum_{k=1}^N J_{jk} \; y_k\, , \quad j=1, \ldots , N \, ,
\end{equation}
to study the stability of large complex systems. Here $J=(J_{jk})$ is the coupling matrix and $\mu >0$.
In the absence of interactions, i.e., when all $J_{jk}=0$,  system (\ref{May1}) is self-regulating: if disturbed from the equilibrium $y_1=y_2=\ldots=y_N=0$ it returns back with some characteristic relaxation time set by $\mu $.
In an ecological context $y_j(t)$ is interpreted as the variation about the equilibrium value, $y_j=0$, in the population density of species $j$ at time $t$. The element $J_{jk}$ of the  coupling matrix $J$, which is known as the community matrix in ecology, measures the per capita effect of species $k$ on species $j$ at the presumed equilibrium.
Generically, the community matrix is asymmetric, $J_{jk}\not =J_{kj}$.

For complex multi-species systems information about the interaction between species is rarely available at the level of detail sufficient for the exact computation of the community matrix and a subsequent stability analysis. Instead, May considered an ensemble of community matrices $J$ assembled at random, whereby the matrix elements $J_{jk}$ are sampled from a probability distribution with zero mean and a prescribed variance $\alpha^2$.
This is similar to the approach taken by Wigner in his description of statistics of energy levels of heavy nuclei via eigenvalues of random matrices, which proved to be very fruitful \cite{Porter65}.  A detailed review of May's model in the light of recent advances in random matrix theory can be found in \cite{AlessinaRev}.

The linear system (\ref{May1}) is stable if, and only if, all the eigenvalues of  $J$ have real parts less than $\mu$.  Invoking Wigner's arguments for studying eigenvalues of large random matrices, May claimed that for large $N$ the largest real part of the eigenvalues of $J$ is typically $\alpha\sqrt{N}$. Obviously,
the model's stability is then controlled by the ratio $m=\mu/(\alpha\sqrt{N})$. For $N$ large,  system (\ref{May1}) will almost certainly be stable if $m>1$ and unstable if $m<1$, with a sharp transition between the two types of behaviour with changing either $\mu$, $\alpha$ or $N$. In particular, for fixed $\mu,\alpha$  system (\ref{May1}) will almost certainly become unstable for $N$ sufficiently large.

Despite the simplistic character of May's model \cite{James2015}, his pioneering work gave rise to a long standing `stability versus diversity' debate, which is not fully settled yet \cite{AlessinaRev,McCann,Boyd,Allesina,Johnson}, and played a fundamental role in theoretical ecology by prompting ecologists to think about special features of real multi-species ecosystems that help such systems to remain stable. Variations of May's model are still being discussed nowadays in the context of neighbourhood stability, see \cite{AlessinaRev}  and references therein.

One obvious limitation of the neighbourhood stability analysis is that it gives no insight into the model behaviour beyond the instability threshold. Hence May's model has only limited bearing on the dynamics of populations operating out-of-equilibrium. An instability does not necessarily imply lack of persistence: populations could coexist thanks to limit cycles or chaotic attractors, which typically originate from unstable equilibrium points. Important questions posing serious challenge  then relate to classification of equilibria by stability, studying their basins of attraction, and other features of global dynamics \cite{AlessinaRev}. Over the last years, as the computing power grew, non-linear models have increasingly been used to investigate population dynamics on the global scale by means of numerical integration of the corresponding system of ODEs \cite{YI92,WM04,DBWM05,RSB14}. Although such investigations captured a rich variety of types of behaviour such as fold bifurcations  when points of equilibrium merge/annihilate \cite{DKG12}, or various types of chaotic dynamics \cite{HP91,McCY94b}, they provide little analytic insight and are limited to small to medium-sized systems.

In this paper we attempt to investigate the generic properties of the \emph{global} dynamics of large complex multi-species systems by retaining in our model only the bare essentials - nonlinearity and stochasticity.  Much in the spirit of May's original approach,  the model we  propose  is simple enough to allow for an analytic treatment yet at the same time is rich enough to exhibit a non-trivial behaviour.  In particular, our model captures an instability transition of the May-Wigner type, but now on the global scale. It also sheds additional light on the nature of this transition by relating it to an exponential explosion in the number of equilibria.  Interestingly,  despite the nonlinear setting of the problem the random matrix ideas again play a central role in our analysis.

 Similar to the May's linear model our toy model is likely to have rather limited practical significance for quantitative description of real ecosystems but it might provide insight into the generic qualitative features of such systems and beyond, e.g. machine learning \cite{GF} or financial ecosystems \cite{HM2011,FS13}. The idea of destabilisation by interaction is of relevance  far beyond the mathematical ecology context as applications of systems of many coupled non-linear ODEs are vast (e.g. complex gene regulatory networks \cite{deJong}, neural networks \cite{SCS,WT}, or random catalytic reaction networks \cite{SFM}).

%------------------------------------------------

\section{Model}

Consider a system of $N$ coupled non-linear autonomous ODEs of the form
 \begin{equation}\label{1}
 \frac{d{x}_i}{dt}=-\mu x_i+f_i(x_1,\ldots, x_N), \quad i=1,\ldots,N,
 \end{equation}
where  $\mu>0$  and  the components $f_{i}({\textbf x})$ of the vector field $\textbf{f}=(f_1,\ldots, f_N)$  are zero mean random functions
of the state vector $\textbf{x}=(x_1,\ldots, x_N)$. To put this model in the context of the discussion above, if $\mathbf{x}_e$ is an equilibrium of (\ref{1}),  i.e., if $-\mu \mathbf{x}_e +\mathbf{f}(\mathbf{x}_e)=\mathbf{0}$
then, in the \emph{immediate} neighbourhood of $\mathbf{x}_e$ system (\ref{1}) reduces to May's model (\ref{May1}) with $\mathbf{y}=\mathbf{x}-\mathbf{x}_e$ and $J_{jk}=(\partial f_j/ \partial x_k) (\mathbf{x}_e)$.

The non-linear system (\ref{1}) may have multiple equilibria whose number and locations depend on the realisation of the random field  $\textbf{f}(\textbf{x})$. To visualise the global picture, it is helpful to consider first a special case of a gradient-descent flow,   characterised by the existence of a potential function $V(\mathbf{x})$ such that $\mathbf{f}=-\nabla V$. In this case, system (\ref{1}) can be rewritten as $ d\mathbf{x}/dt=-\nabla L$, with $L(\mathbf{x})=\mu{|{\bf x}|^2}/{2}+V({\bf x})$ being the associated Lyapunov function describing the effective landscape. In the domain of $L$, the state vector $\mathbf{x}(t)$ moves in the direction of the steepest descent, i.e., perpendicular to the level surfaces $L(\mathbf{x})=h$  towards ever smaller values of $h$. This provides a useful geometric intuition. The term $\mu{|{\bf x}|^2}/{2}$ represents the globally confining parabolic potential, i.e., a deep well on the surface of $L(\mathbf{x})$, which does not allow  $\mathbf{x}$ to escape to infinity. At the same time the random potential $V(\mathbf{x})$ may generate many local minima of $L(\textbf{x})$ (shallow wells) which will play the role of attractors for our dynamical system.  Moreover, if the  confining term is strong enough then the full landscape will only be a small perturbation of the parabolic well, typically with a single stable equilibrium  located close to ${\mathbf x}=0$. In the opposite case of relatively weak confining term,  the disorder-dominated landscape will be characterised by a complicated random topology with many points of equilibria,  both stable and unstable. Note that in physics, complicated energy landscapes is a generic feature of glassy systems with  intriguingly slow long-time relaxation and non-equilibrium dynamics, see e.g. \cite{APRV2000}.

The above picture of a gradient-descent flow is however only a very special case since the generic systems of ODEs (\ref{1}) are not gradient. The latter point can easily be understood in the context of model ecosystems. For, by linearising a gradient flow in a vicinity of any equilibrium, one always obtains a symmetric community matrix, whilst the community matrices of model ecosystems are in general asymmetric.  Note also a discussion of an interplay between non-gradient dynamics in random environment and glassy behaviour in \cite{CKDP}.

To allow for a suitable level of generality we therefore suggest to choose the $N-$dimensional vector field $\mathbf{f}(\mathbf{x})$  as a sum of `gradient' and non-gradient (`solenoidal') contributions:
\begin{equation}\label{2}
  f_i({\bf x})=-\frac{\partial V({\bf x})}{\partial x_i}+\frac{1}{\sqrt{N}}\sum_{j=1}^N\frac{\partial A_{ij}({\bf x})}{\partial x_j}, \quad i=1,\ldots,N,
 \end{equation}
where we require the matrix $A({\bf x})$ to be antisymmetric: $A_{ij}=-A_{ji}$. The meaning of this decomposition is that vector fields can be generically divided into a conservative irrotational component, sometimes called `longitudinal', whose gradient connects the attractors or repellers and a solenoidal curl field, also called `transversal'. As discussed in, e.g.,  \cite{Aurell} such a representation is closely related to the so-called Hodge decomposition of differential forms and generalises the well-known Helmholtz decomposition of the $3-$dimensional vector fields into curl-free and divergence-free parts to higher dimensions. Correspondingly, we will call $V({\bf x})$ the scalar potential and the matrix $A({\bf x})$ the vector potential. The normalising factor $1/\sqrt{N}$ in front of the sum on the right-hand side in (\ref{2}) ensures that the transversal and longitudinal parts  of $\mathbf{f}({\bf x})$ are of the same order of magnitude for large $N$.

Finally, to make the model as simple as possible and amenable to  a  rigorous and detailed  mathematical analysis we choose the scalar potential $V({\bf x})$ and the components $A_{ij}({\bf x})$, $i<j$, of the vector potential to be statistically independent, zero mean Gaussian random fields, with  smooth realisations and the additional assumptions of  {\it homogeneity} (translational invariance) and {\it isotropy} reflected in the covariance structure:
\begin{eqnarray}\label{3a}
\langle V({\bf x}) V({\bf y})\rangle&\!\!\!=\!\!\!&v^2 \Gamma_V\! \left(|{\bf x}-{\bf y}|^2 \right); \\
\label{3b}
\langle A_{ij}({\bf x}) A_{nm}({\bf y})\rangle&\!\!\!=\!\!\!&a^2 \Gamma_A\! \left(|{\bf x}-{\bf y}|^2\right) \! \left(\delta_{in}\delta_{jm}-\delta_{im}\delta_{jn}\right).
\end{eqnarray}
Here the angular brackets $\langle ... \rangle$ stand for the ensemble average over all realisations of $V(\mathbf{x})$ and $A(\mathbf{x})$, and $\delta_{in}$ is the Kronecker delta: $\delta_{in}=1$ if $i=n$ and zero otherwise.

For simplicity, we also assume that the functions $\Gamma_V(r)$ and $\Gamma_A(r)$ do not depend on $N$.  This implies \cite{Sh38}
\[
\Gamma_{\sigma} (r)=\int_0^{\infty} \exp ({- s\, r}) \gamma_{\sigma}(s) ds, \quad \sigma=A,V,
\]
where the `radial spectral' densities $\gamma_{\sigma}(s) \ge 0$  have finite total mass: $\int_0^{\infty} \gamma_{\sigma}(s) ds < \infty$. We normalize these densities by requiring that $\Gamma^{\prime\prime}_{\sigma}(0) = \int_0^{\infty} s^2 \gamma_{\sigma}(s) ds=1$.
The ratio
\[
\tau={v^2}/{(v^2+a^2)}, \quad 0\le \tau \le 1,
\]
is a dimensionless measure of the relative strengths of the longitudinal and  transversal
%the two
components of $\mathbf{f}(\textbf{x})$: if $\tau=0$ then $\mathbf{f}(\textbf{x})$ is divergence free and if $\tau=1$ it is curl free.

\section{Results}
 Determining and classifying all points of equilibria of a dynamical system with many degrees of freedom is a well-known formidable analytical and computational problem. In this paper we shall focus our investigation on the simplest, yet informative characteristic of system (\ref{1}) by counting its total number of equilibria, that is the total number ${\cal N}_{tot}$ of solutions of the  simultaneous equations
\begin{equation}\label{b1}
-\mu x_i+f_i(x_1,\ldots, x_N)=0, \, i=1,\ldots, N.
\end{equation}
Certainly, finding ${\cal N}_{tot}$ is a good starting point of any phase portrait analysis.

Had we restricted ourselves to the gradient-descent flows, ${\cal N}_{tot}$  would simply count the number of stationary points  (minima, maxima, or saddle-points) on the surface of the Lyapunov function $L({\bf x})$. The problem of counting and classifying stationary points of high-dimensional random energy landscapes of various types attracted considerable interest in recent years \cite{my2004,FyoNad,Auf1,Nicolaescu2,my2014,Subag2015}. In particular, works \cite{my2004,FyoNad} study such energy landscapes generated by a potential  equivalent to the above Lyapunov function.  One of the main  conclusions of that study is that for $N$ large the topology of the Lyapunov function changes drastically with decrease
of the strength of the confining term relative to that of the interaction term in $L(\mathbf{x})$. The change manifests itself in the emergence of multitude of equilibria, exponential in number. Such a transition is intimately connected to the spin-glass like restructuring of the Boltzmann-Gibbs measure induced by the Lyapunov function when the latter is treated as an effective energy landscape.

We shall prove below that for $N$ large the general autonomous system (\ref{1})--(\ref{2}) exhibits a similar drastic change in the total number of equilibria
when the control parameter
\[
m=\frac{\mu}{\alpha \sqrt{N}}, \quad \text{where $\alpha = 2 \sqrt{v^2+a^2}$},
\]
drops below the threshold  value $m_c=1$. As in the case of gradient systems, the proof involves the Kac-Rice formula as a starting point. However, performing the subsequent steps requires  quite different mathematical techniques due to the  asymmetry of the Jacobian matrix for non-gradient systems.

The Kac-Rice formula, see e.g. \cite{math2}, counts solutions of simultaneous algebraic equations. Under our assumptions (homogeneity, isotropy and Gaussianity of $V$ and $A$), this formula yields the ensemble average of ${\cal N}_{tot}$ in terms of that of the modulus of the spectral determinant of the Jacobian matrix $(J_{ij})_{i,j=1}^N$,  $J_{ij}={\partial f_i}/{\partial x_j}$ (see Materials and Methods):
\begin{equation}\label{5}
\left\langle{\cal N}_{tot}\right\rangle=\frac{1}{\mu^N}\left\langle\left| \det\left( -\mu\delta_{ij} +J_{ij}\right)\right|\right\rangle,
\end{equation}
thus bringing the original non-linear problem into the realms of the random matrix theory.

The probability (ensemble) distribution of the matrix $J$ can easily  be determined in closed form. Indeed, the  matrix entries of $J$ are zero mean Gaussian variables and their covariance structure, at spatial point $\textbf{x}$, can be obtained from  (\ref{3a})--(\ref{3b}) by differentiation:
%\begin{equation}%\label{6a}
\[
\left\langle J_{ij}J_{nm} \right\rangle=\alpha^2[ (1+\epsilon_N)\delta_{in}\delta_{jm}+  (\tau - \epsilon_N) (\delta_{jn}\delta_{im}+\delta_{ij} \delta_{mn} )]\, ,
\]
%\end{equation}
where $\epsilon_N=(1-\tau)/N$. Thus, to leading order in the limit $N\to\infty$,
\begin{equation}\label{JX}
J_{ij}=\alpha (X_{ij}+ \sqrt{\tau}\delta_{ij}\xi),
\end{equation}
where $X_{ij}$, $i,j=1, \ldots, N$ are zero mean Gaussians with
\begin{equation}\label{X}
\left\langle X_{ij}X_{nm} \right\rangle=\delta_{in}\delta_{jm}+  \tau \delta_{jn}\delta_{im}\, ,
\end{equation}
and $\xi$ is a standard Gaussian, $\xi \sim N(0,1)$, which is statistically independent of $X=(X_{ij})$.
Note that for the divergence free fields $\mathbf{f}(\mathbf{x})$ (i.e., if $\tau=0$) the entries of  $J$ are statistically independent in the limit $N\to\infty$, exactly as in May's model. On the other side, if  $\mathbf{f}(\mathbf{x})$ has a longitudinal component ($\tau >0$) then this implies positive correlation between the pairs of matrix entries of $J$ symmetric about the main diagonal: $\left\langle X_{ij}X_{ji} \right\rangle = \tau$ if $i\not= j$. Such distributions of the community matrix has also been used in the neighbourhood  stability analysis of model ecosystems \cite{Allesina}. Finally, in the limiting case of curl free fields ($\tau=1$), the matrix $J$ is real symmetric.

The representation (\ref{JX}) comes in handy as it allows one to express (\ref{5}) as a random matrix integral:
\begin{equation}\label{8}
\left\langle{\cal N}_{tot}\right\rangle= \frac{N^{-\frac{N}{2}}}{m^N} \int_{-\infty}^{\infty} \left\langle | \det (x\, \delta_{ij}-X_{ij})|\right\rangle_{X_N}\frac{e^{-\frac{Nt^2}{2}} dt}{\sqrt{2\pi/N}},
\end{equation}
 where  $x=\sqrt{N}(m+t\sqrt{\tau})$ and the angle brackets $\langle \ldots \rangle_{X_N}$ stand for averaging over the real elliptic ensemble of random $N\times N$ matrices $X$ defined in (\ref{X}), see also (\ref{JPD}).
This one-parameter family of random matrices interpolates between the Gaussian Orthogonal Ensemble of real symmetric matrices (GOE, $\tau=1$) and real Ginibre ensemble of fully asymmetric matrices (rGinE, $\tau=0$), see \cite{KS} for discussions. Both rGinE and its one-parameter extension (\ref{X}) have enjoyed considerable interest in the literature in recent years \cite{AkKanz2007,SW2008,BS2009,ForNag2008,AP2014}.

The matrix $X$ is  asymmetric (unless $\tau =1$) and can have real as well as complex eigenvalues. The latter come in complex-conjugate pairs. Their density, in the limit  $N\to\infty$, is constant inside the ellipse with the main half-axis $\sqrt{N}(1\pm \tau)$ and vanishes sharply outside \cite{Som88,ForNag2008,KS}. The corresponding theorem is known as the Elliptic Law and its validity extends beyond the Gaussian matrix distributions \cite{G85,Nguyen14}. However, in the context of our investigation it is the density of real eigenvalues of $X$ that appears to be most relevant.

Denote by $\rho_{N}^{(r)}(x)$ the density of real eigenvalues of $N\times N$ matrices $X$ (\ref{X}) averaged over all realisations of $X$. It is convenient to normalize $\rho_{N}^{(r)}(x)$  in such a way that  $\int_{\alpha}^{\beta} \rho_{N}(x)\,dx$ gives the average number of real eigenvalues of $X$ in the interval $[\alpha,\beta]$.  A crucial observation is that
$\rho_{N}^{(r)}(x)$  is directly related to the averaged value of the modulus of the determinant that appears in (\ref{8}). Namely,
\begin{equation}\label{9}
\left\langle | \det (x\delta_{ij}-X_{ij})|\right\rangle_{X_N} ={\cal C}_N(\tau)\,  e^{\frac{x^2}{2(1+\tau)}}\, \rho^{(r)}_{N+1}(x)\, ,
\end{equation}
where ${\cal C}_N(\tau)=2\sqrt{1+\tau}\,  (N-1)!/(N-2)!!$ and  $\rho^{(r)}_{N+1}(x)$ is the average density of real eigenvalues of matrices $X$ of size  $(N+1)\times (N+1)$. For the limiting case $\tau=0$ this relation appeared originally in \cite{EKS}, and it can be extended to any $\tau\in [0,1)$ without much difficulty  (see SI for a derivation of (\ref{9}) following the approach of \cite{FK2007} ). In the limiting case of real symmetric matrices $\tau=1$, all eigenvalues of $X$ are real and relation (\ref{9}) is also valid \cite{my2004}.

Combining  (\ref{8}) and (\ref{9}) and changing the variable of integration from $t$ to  $\lambda=m+t\sqrt{\tau}$, one can express $\left\langle{\cal N}_{tot}\right\rangle$ for system (\ref{1}) with $N$ degrees of freedom in terms of the density of real eigenvalues in the elliptic ensemble of random matrices (\ref{X}) of size $(N+1)\times (N+1)$:
\begin{equation}\label{11}
\left\langle{\cal N}_{tot}\right\rangle= \frac{{\cal K}_N(\tau)}{m^N}\int_{-\infty}^{\infty}
\!\! e^{-N S(\lambda)} \rho_{N+1}(\lambda\sqrt{N})\, \frac{d\lambda}{\sqrt{2\pi}},
\end{equation}
where $S(\lambda)=\frac{(\lambda-m)^2}{2\tau} - \frac{\lambda^2}{2(1+\tau)}$ and ${\cal K}_N(\tau)={N^{\frac{-N+1}{2}}{\cal C}_N(\tau)}/\!{\sqrt{\tau}}$. The importance of this relation is due to the fact that $\rho_{N}^{(r)}(x)$ is known  in closed form in terms of Hermite polynomials \cite{ForNag2008}. This allows us to carry out an asymptotic evaluation of the integral in (\ref{11}) and calculate $\left\langle{\cal N}_{tot}\right\rangle $ in the limit $N\to \infty$. The key finding that emerges from this calculation is that $\left\langle{\cal N}_{tot}\right\rangle $ changes drastically around $m=1$.
If $m>1$ then
\begin{equation}\label{13a}
 \lim_{N\to\infty}\left\langle {\cal N}_{tot} \right\rangle = 1.
\end{equation}
On the other hand, if $0<m<1$ then, to leading order in the limit $N\to \infty$,
\begin{equation} \label{13b}
\left\langle {\cal N}_{tot} \right\rangle \, = \, \gamma_{\tau} \,
e^{ N\Sigma_{tot}(m) }\,,
\end{equation}
where $ \Sigma_{tot}(m)=\frac{1}{2}(m^2-1)-\ln{m}>0$  for all  $0<m<1$. Therefore,  if $m<1$ then $\left\langle {\cal N}_{tot} \right\rangle$ grows exponentially with $N$.
The factor in front of the exponential in (\ref{13b})  is given  by $\gamma_{\tau}= \sqrt{{2(1+\tau)}/{(1-\tau)}}$ as long as $\tau<1$.
The gradient limit $\tau=1$ can be approached by scaling $\tau$ with $N$. Setting $\tau=1-\frac{u^2}{N}, \, 0\le u<\infty$, one obtains
$ \gamma_{\tau}= 4\sqrt{\frac{N}{\pi}}\,\int_0^{\sqrt{1-m^2}}\,e^{-u^2p^2} dp$.
This regime describes a weakly non-gradient flow.  The corresponding regime for ensembles of asymmetric matrices was discovered long ago \cite{FyoKhorSom1997,Efe1997}.

Close to the phase transition point $m=1$ the complexity exponent vanishes quadratically, $\Sigma_{tot}=(1-m)^2$ as $m\to1$,
implying that the width of the transition region around $m=1$ is $1/\sqrt{N}$. According to the general lore of phase transitions, for large but finite $N$ there exists a `critical regime' $m=1+\kappa N^{-1/2}$ where the number of equilibria changes smoothly between the two `phases' (\ref{13a}) and  (\ref{13b}). A quick inspection of (\ref{11}) shows that the corresponding crossover profile is determined by the profile of $\rho_{N}^{(r)}(x)$ in the vicinity of the `spectral edge' $x=(1+\tau)\sqrt{N}$, see Materials and Methods.  After rescaling $\lambda$, $\lambda=1+\tau+\frac{\zeta\sqrt{1-\tau^2}}{\sqrt{N}}$, the density  $\rho_{N}^{(r)}(\lambda\sqrt{N})$ converges to $ \frac{1}{\sqrt{1-\tau^2}}\rho_{edge}^{(r)}(\zeta)$ in the limit $N\to\infty$, where \cite{ForNag2008}:
\begin{equation}\label{15}
\rho_{edge}^{(r)}(\zeta)=\frac{1}{2\sqrt{2\pi}}\left\{
\erfc (\sqrt{2}\zeta) +\frac{1}{\sqrt{2}}e^{-\zeta^2}\left[1+\erf\left(\zeta\right)\right]\right\}
\end{equation}
with $\erf(x)=1-\erfc(x)= \frac{2}{\sqrt{\pi}}\int_0^xe^{-t^2}\,dt$. In terms of $\rho_{edge}^{(r)}(\zeta)$, the critical crossover profile is given by
\begin{equation}\label{16}
\left\langle{\cal N}_{tot}\right\rangle = \gamma_{\tau}\,e^{\kappa^2} \!\!\!
 \int_{-\infty}^{\infty}e^{-\frac{t^2}{2}}\,\rho_{edge}^{(r)} \left(c_{\tau} t+\kappa \, \frac{\gamma_{\tau}}{\sqrt{2}}  \right)\, dt
\end{equation}
 where $c_{\tau} =\sqrt{\tau/(1-\tau)}$. The right-hand side of (\ref{16})  interpolates  smoothly between the two regimes  (\ref{13a}) and (\ref{13b}),  when parameter $\kappa$ runs from $\kappa= -\infty$ to  $\kappa = +\infty$.

Although our investigation is concerned with the ensemble average of the number of equilibria ${\cal N}_{tot}$, we expect that in the limit $N\to \infty$ the deviations of ${\cal N}_{tot}$ from its average $\langle {\cal N}_{tot} \rangle$ are relatively small.  This is certainly the case above the critical threshold, for $m>1$. For,
under some additional technical assumptions on the decay of correlations for $\mathbf{f}(\mathbf{x})$, the system  (\ref{1}) will almost certainly have at least one stationary point, see \cite{Pit} for the relevant results about the maxima of homogeneous Gaussian fields.  Therefore,  ${\cal N}_{tot}\ge 1$ and the established convergence of $\langle {\cal N}_{tot} \rangle$ to 1 in the limit $N\to\infty$ actually implies that the probability for  ${\cal N}_{tot}$ to take other values than one is close to zero for large $N$. The problem of estimating the deviation of   ${\cal N}_{tot}$ from its average value in the opposite regime $0<m<1$ is much harder and is an open and interesting question\footnote{In this context we  would like to mention the recent work of Subag \cite{Subag2015} who proved that the deviations of $ {\cal N}_{tot}$ from $\langle {\cal N}_{tot} \rangle$ in the spherical $p$-spin model are negligible in the limit of large system size. Though that  model is different from ours, it is not dissimilar to the gradient limit of $\tau=1$ of our model \cite{my2014}, for instance, the average number of equilibria grows exponentially with $N$ \cite{Auf1}. Thus one might hope to adopt the technique of \cite{Subag2015} to our model. Another relevant reference is \cite{Nicolaescu3}.}.

%------------------------------------------------

\section{Discussion}

In this paper we introduced a model describing generic large complex systems and examined the dependence of the total number of equilibria in such systems on the system complexity as measured by the number of degrees of freedom and the interaction strength.  The inspiration for our work came from May's pioneering study \cite{May1972} of the \emph{neighbourhood} stability of large model ecosystems. Our outlook is complementary to that of May's in that it adopts a \emph{global} point of view which is not limited to the neighbourhood of the presumed equilibrium.

In the context of model ecosystems our analysis is applicable to complex multi-species communities in which each kind of species on its own becomes extinct and thus interaction between species is key to persistence of the community. The key feature of our analysis is that in the presence of interactions, as the complexity increases, there is an abrupt change from a simple set of equilibria (typically, a single equilibrium for large number of species $N\gg 1$) to a complex set of equilibria, with their total number growing exponentially with $N$. In the latter regime,  we expect the stable equilibria to be only a tiny proportion of all the multitude of equilibria, see discussion below, which is indicative of long relaxation times and transient non-equilibrium behaviour.

We expect this sharp transition in the phase portrait to be shared by other systems of randomly coupled autonomous ODE's  with large number of degrees of freedom.  To that end, it is appropriate to mention that very recently a similar `explosion in complexity' was reported in a model of neural network consisting of randomly interconnected neural units \cite{WT}.
The model considered in \cite{WT} is essentially of the form (\ref{1})
but with the particular choice of $f_i=\sum_{j}J_{ij}S(x_j)$ where $S$ is an odd sigmoid function representing the synaptic nonlinearity and $J_{ij}$ are independent centred Gaussian variables representing the synaptic connectivity between neuron $i$ and $j$.  Although being Gaussian,  the corresponding (non-gradient) vector field is not homogeneous and thus seems rather different from our choice and not easily amenable to a rigorous analysis.   Nevertheless, a shrewd semi-heuristic analysis of \cite{WT} revealed that  close to the critical coupling threshold the two models actually display very similar behaviour, described essentially by the same exponential growth in the total number of equilibria  with rate $\Sigma_{tot}(m)$. This fact points towards considerable universality of the transition from (\ref{13a}) to (\ref{13b}) and suggests that the crossover function (\ref{16}) may  be universal as well.

Our model captures an abrupt change %in the global phase portrait
in the dynamics of large complex systems on the macroscopic scale. %complexity increases.
At the same time zooming in to classify each and every equilibrium point into locally stable or unstable seems a hard task. For, although linearising the field  $\mathbf{f}(\mathbf{x})$ around a given equilibrium is fairly straightforward, with the outcome being the Jacobian matrix (\ref{JX}),
conditioning by the positions of equilibria and taking into account all eventualities is a highly non-trivial task. Given the stochastic setup of our model the question about stability of individual equilibria  may be even the wrong question to ask, whereas addressing the statistics of the {\it number} of stable equilibria seems very appropriate.

Arguments similar to those in the previous section yield the ensemble average of the total number of stable equilibria, $\langle {\cal N}_{st} \rangle$, over all realisations of the vector field $\mathbf{f}(\mathbf{x})$ in terms the random matrix integral (cf.(\ref{8})):
\begin{equation}\label{8new}
\left\langle{\cal N}_{st}\right\rangle\!=\! \frac{N^{-\frac{N}{2}}}{m^N} \int\limits_{-\infty}^{\infty} \!\! \left\langle\det (x\, \delta_{ij}-X_{ij}) \chi_x (X_{ij}) \right\rangle_{X_N}  \frac{e^{-\frac{Nt^2}{2}} dt}{\sqrt{2\pi/N}}\, ,
\end{equation}
where $\chi_x(X)=1$, if all $N$ eigenvalues of matrix $X$ have real parts less than the spectral parameter $x=\sqrt{N}(m+t\sqrt{\tau})$, and  $\chi_x(X)=0$ otherwise. In the limiting case of a purely gradient dynamics $\tau=1$, the rescaled Jacobian matrix $X$ is real symmetric with all $N$ eigenvalues real. In this case the above integral can be related to the probability density of the maximal eigenvalue of the GOE matrix \cite{FyoNad,Auf1}, with the latter being a well-studied object in the random matrix theory, see e.g. \cite{MajSch2014} and references therein. This observation can then be used to evaluate $\left\langle{\cal N}_{st}\right\rangle$ for $N\gg 1$. One finds \cite{FyoNad} that  $\langle {\cal N}_{st} \rangle \to 1$ if $m>1$, whilst if $0<m<1$ then, to leading order in $N$,   $\langle {\cal N}_{st}\rangle \propto e^{N\Sigma_{st}}$, with $0< \Sigma_{st}<\Sigma_{tot}$.
Thus, in the case of purely gradient dynamics, as the complexity increases, large nonlinear autonomous systems assembled at random undergo an abrupt change from a typical phase portrait with a single stable equilibrium to a phase portrait dominated by an exponential number of unstable equilibria with an admixture of a smaller, but still exponential in $N$, number of stable equilibria.

It was suggested to us by J.-P. Bouchaud that in the general case of non-gradient dynamics $0\le \tau<1$,  it would be natural to expect a further phase transition in the plane $(m, \tau) $ such that below a certain number $\tau_c(m)$ stable equilibria are no longer exponentially abundant  in the limit $N\to \infty$ (i.e. $\Sigma_{st}(m,\tau)\to 0$), with further implications for the global dynamics.
Unfortunately, for a fixed $0\le \tau<1$ only vanishing fraction of order of $N^{-1/2}$ of eigenvalues of $X$ remain real, and the relation of the integral in Eq. (\ref{8new}) to statistics of the largest {\it real} eigenvalue in the elliptic ensemble seems to be lost. This fact prevented us so far from reliable counting of stable equilibria for the general case of non-gradient flows. In principle, for given values of parameters $N,\tau,m$ one may attempt to evaluate the ensemble average  in the integral in Eq.   (\ref{8new}) numerically, and then to evaluate numerically the integral itself. Although such a  procedure seems tractable, its actual implementation with sufficient precision is not straightforward, especially in the limit $N\to \infty$ due to the exponentially large values involved.
Clarification of the status of the picture suggested by J.-P. Bouchaud and related studies remain an  important  outstanding issue and is left for a future work.

%----------------------------------------------------------------------------------------
%	MATERIALS AND METHODS
%----------------------------------------------------------------------------------------

%% Optional Materials and Methods Section
%% The Materials and Methods section header will be added automatically.

\begin{materials}\fontsize{8}{9}\selectfont\materialfont
\textbf{Kac-Rice Formula}\quad
The expected number $\langle {\cal N}_{tot}\rangle$ of simultaneous solutions to the system of equations  (\ref{b1})  in $\mathbb{R}^N$ is given by the formula, see e.g. \cite{math2},
 \begin{equation}\label{4}
 %{\cal N}_{tot}
\langle {\cal N}_{tot}\rangle \!=\!\int_{\mathbb{R}^N} \!\! \langle \delta\left(-\mu \mathbf{x}\!+\! \mathbf{f}(\mathbf{x})\right)\left| \det\left( -\mu\delta_{ij}\!+\!
 J_{ij}(\mathbf{x}) \right) \right|\rangle \,dx^N\, ,
 \end{equation}
where $\delta(\mathbf{x})$ is the multivariate Dirac $\delta$-function, $dx^N$ is the volume element in $\mathbb{R}^N$ and $J_{ij}(x) ={\partial f_i}/{\partial x_j}$ are matrix elements of the Jacobian matrix $J=(J_{ij})$. By our assumptions (\ref{3a}) -- (\ref{3b}) the random field $\mathbf{f}(\mathbf{x})$ is homogeneous and isotropic. For such fields samples of $\mathbf{f}$ and $J$  taken in one and the same spatial point $\mathbf{x}$ are uncorrelated, $\langle f_l \cdot {\partial f_i}/{\partial x_j} \rangle=0$ for all $i,j,l$.  This is well known and can be checked by straightforward differentiation. In addition, the field $\mathbf{f}$ is Gaussian, hence the $\mathbf{f}(\mathbf{x})$ and $J(\mathbf{x})$ are actually statistically independent. This simplifies the evaluation of the integral in (\ref{4}) considerably. Indeed, the statistical average in (\ref{4}) factorizes and, and since  $\langle |\det\left( -\mu\delta_{ij}+J_{ij}(x) \right) | \rangle$ does not vary with $\mathbf{x}$, the integrand effectively reduces to
 \begin{equation}\label{a1}
\langle \delta ((-\mu \mathbf{x} +\mathbf{f}(\mathbf{x}))\rangle =\int \frac{d k^N}{(2\pi)^N}\,  e^{-\mu \mathbf{k}\cdot \mathbf{x}} \,
\langle e^{i\mathbf{k}\cdot \mathbf{f}(\mathbf{x}) }\rangle .
\end{equation}
Furthermore, at every spatial point $\mathbf{x}$ the vector $\mathbf{f}(\mathbf{x})$ is Gaussian with uncorrelated and identically distributed components,
\[
\langle f_i(\mathbf{x})f_j(\mathbf{x})\rangle =\delta_{ij} \sigma^2, \quad \sigma^2=2v^2|\Gamma^{\prime}_V(0)|+2a^2  |\Gamma^{\prime}_A(0)|\frac{N-1}{N}.
\]
Therefore $\langle e^{i\mathbf{k}\cdot \mathbf{f}(\mathbf{x}) }\rangle = e^{- \sigma^2 |\mathbf{k}|^2/2}$, and
evaluating the integral on the right-hand side in (\ref{a1}) one arrives at (\ref{5}).

\noindent \textbf{Real elliptic matrices and asymptotics of $\langle  {\cal N} \rangle_{tot}$} \quad The joint probability density function ${\cal P}_N(X)$ of the matrix entries in the elliptic ensemble of real Gaussian random matrices $X$ of size $N\times N$ is given by
\begin{equation}\label{JPD}
{\cal P}_N(X)={\cal Z}_N^{-1}\exp\Big[-\frac{1}{2(1-\tau^2)}\Tr \big( X X^T-\tau X^2\big) \Big],
\end{equation}
where ${\cal Z}_N$ is the normalisation constant and $\tau\in[0,1)$. It is straightforward to verify that the covariance of  matrix entries $X_{ij}$ is given by the expression specified in (\ref{X}).
The mean density of real eigenvalues of $\rho_{N}^{(r)}(x)$ in the elliptic ensemble (\ref{JPD}) is known in closed form in terms of Hermite polynomials, see \cite{ForNag2008}. Assuming for simplicity that $N+1$ is even,
 one has $\rho_{N+1}^{(r)}(x)=\rho_{N+1}^{(r),1}(x)+\rho_{N+1}^{(r),2}(x)$ where
\begin{equation}\label{10a} \rho_{N+1}^{(r),1}(x)=
\frac{1}{\sqrt{2\pi}} \, \sum_{k=0}^{N-1}\, \frac{\big|\psi^{(\tau)}_k(x)\big|^2}{k!},
\end{equation}
and
\begin{equation}\label{10b}
 \rho_{N+1}^{(r),2}(x)=
\frac{1}{\sqrt{2\pi}(1+\tau)(N-1)!}\,  \psi^{(\tau)}_{N}(x) \int_0^{x}\psi^{(\tau)}_{N-1}(u)\,du.
\end{equation}
Here $\psi^{(\tau)}_k(x)=e^{-\frac{x^2}{2(1+\tau)}}h^{(\tau)}_k(x)$ and $h^{(\tau)}_k(x)$ are rescaled Hermite polynomials,
$h^{(\tau)}_{k}(x)=\frac{1}{\sqrt{\pi}}\int_{-\infty}^{\infty}e^{-t^2}
\left(x+it\sqrt{2\tau}\right)^k\,dt$.
This, together with the integral (\ref{11}) allow one to evaluate $\langle  {\cal N} \rangle_{tot}$ in the limit $N\to\infty$.
We shall sketch the corresponding evaluation below.

The asymptotics of $\rho_{N}^{(r)}(x)$ in the bulk and at the edge of the support of the distribution of real eigenvalues in the real elliptic ensemble were found in \cite{ForNag2008}, and outside the support it can also be readily extracted using (\ref{10a})-(\ref{10b}). In particular,  in the bulk, i.e., for  $|x|<(1+\tau)\sqrt{N}$, the contribution of (\ref{10a}) to $\rho_{N}^{(r)}(x)$  is dominant and, to leading order in $N$,
\begin{equation}\label{12a}
\left. \rho_{N+1}^{(r)}(\lambda\sqrt{N})\right|_{|\lambda|<1+\tau}%\approx
=\frac{1}{\sqrt{2\pi(1-\tau^2)}}.
\end{equation}
At the same time for $|x|>(1+\tau)\sqrt{N}$ both (\ref{10a}) and (\ref{10b}) yield exponentially small contributions to $\rho_{N+1}^{(r)}(x)$, with  (\ref{10b}) being dominant. Our evaluation yields
 \[
\left.  \rho_{N+1}^{(r)}(\lambda\sqrt{N}) \right|_{\lambda>(1+\tau)}= Q(\lambda) \exp{-N\Psi(\lambda)}, \quad \mbox{where}\]
 \begin{equation}\label{12b1}
Q(\lambda)=\left[\frac{\tau}{2\pi(1+\tau)}\frac{1}{\sqrt{\lambda^2-4\tau})(\lambda-\sqrt{\lambda^2-4\tau})}\right]^{1/2}\, ,
\end{equation}
 \begin{equation}\label{12b2}
\Psi(\lambda)=\frac{\lambda^2}{2(1+\tau)}-\frac{1}{8\tau}(\lambda-\sqrt{\lambda^2-4\tau})^2-
\ln{\frac{\lambda+\sqrt{\lambda^2-4\tau}}{2\sqrt{\tau}}}\, .
\end{equation}
The form of (\ref{11}) suggest the application of the Laplace method. One easily finds that $S(\lambda)$ has a minimum at $\lambda_*=m(1+\tau)$ which belongs to the domain $|\lambda|<1+\tau$ as long as $0<m<1$.Thus, applying the Laplace method and taking into account the asymptotic formula
\begin{equation}\label{K}
{\cal K}_N(\tau)\approx 2\sqrt{\pi N}e^{-N/2}\sqrt{{(1+\tau)}/{\tau}},
\end{equation}
one arrives at the asymptotic expression (\ref{13b}) for  $\langle {\cal N} \rangle_{tot}$ in the parameter range $0<m<1$. For $m>1$ the saddle-point occurs in the domain $\lambda>1+\tau$ so that the analysis requires search for the minimum of $S(\lambda)+\Psi(\lambda)$. After straighforward algebra we find
$
\frac{d}{d\lambda}\left(S(\lambda)+\Psi(\lambda)\right)=\frac{1}{2\tau}(\lambda+\sqrt{\lambda^2-4\tau})-\frac{m}{\tau}
$
which is equal to zero at $\lambda=\lambda_*=m+\frac{\tau}{m}>1+\tau$. One also verifies that this is a point of minimum for $S(\lambda)+\Psi(\lambda)$ and a further simple calculation then yields $S(\lambda_*)+\Psi(\lambda_*)=-\ln{\left(m/\sqrt{\tau}\right)}$.  Calculating the saddle-point contribution then yields (\ref{13a}).

The above asymptotic analysis assumes that $0\le \tau<1$.  Let us now discuss the modifications required to study the scaling regime of weakly non-gradient flow $\tau\to 1$ for $0<m<1$. We only need to evaluate the leading contribution to $ \rho_{N+1}^{(r)}$ given by (\ref{10a}). By making use of the above integral representation for the scaled Hermite polynomials $h_k^{(\tau)}(x)$ and applying the scaling $\tau=1-\frac{u^2}{N}$, we can write
\[
\rho_{N+1}^{(r),1}(\lambda\sqrt{N})=
\frac{1}{\sqrt{2\pi}}\,\frac{N}{2\pi\tau}\, e^{-N\frac{\lambda^2}{1+\tau}+N\frac{\lambda^2}{\tau}}I_{N+1}(\lambda),
\]
with $I_{N}(\lambda)$ given by
 \[
I_N(\lambda)=\sqrt{\frac{\pi}{N}}e^{-N\frac{\lambda^2}{2}}\int_{-\infty}^{\infty}e^{-\frac{u^2p^2}{2}}
\Phi_{N-2}\left(\frac{N}{2}(p^2+\frac{\lambda^2}{2})\right),
 \]
where  $\Phi_{N}(a)=e^{-a}\sum_{k=0}^{N}{a^k}/{k!}$. Recalling that  the limit of $\Phi_{N}(a)$ as $N\to\infty$ is 1 if $0<a<1$ and 0 if $a>1$, one obtains
\[
\rho_{N+1}^{(r)}(\lambda\sqrt{N})=\frac{1}{\pi}\sqrt{N}\int_0^{\sqrt{1-\frac{\lambda^2}{4}}}e^{-u^2p^2}\,dp.
\]
Substituting this expression into the integrand of (\ref{11}) and evaluating the integral in the limit $N\to \infty$ (hence, $\tau\to 1$) by the Laplace method then yields  $\langle {\cal N} \rangle_{tot}$ in the weakly non-gradient regime.

Finally, our calculation of the profile of $\langle {\cal N} \rangle_{tot}$ in the transitional region $m=1+\kappa N^{-1/2}$ uses the fact that in such a regime the main contribution to the integral (\ref{11}) comes from the neighbourhood of the spectral edge, $\lambda=1+\tau+\frac{\zeta\sqrt{1-\tau^2}}{\sqrt{N}}$, where we have,  to the leading order in $N$,
\[
\frac{e^{-N\left(S(\lambda)+\frac{1}{2}\right)}}{m^N}=\exp{\left[-\frac{1-\tau}{2\tau}(\kappa^2+\zeta^2)
+\frac{\sqrt{1-\tau^2}}{\tau}\kappa\zeta\right]}.
\]
This together with (\ref{K}) and (\ref{15}) converts (\ref{11}) to (\ref{16}).

\end{materials}

%----------------------------------------------------------------------------------------
%	APPENDICES (OPTIONAL)
%----------------------------------------------------------------------------------------

%

%----------------------------------------------------------------------------------------
%	ACKNOWLEDGEMENTS
%----------------------------------------------------------------------------------------

\begin{acknowledgments}
The research of the first author was supported by EPSRC grant EP/J002763/1 ``Insights into Disordered Landscapes via Random Matrix Theory and Statistical Mechanics''. The second author
would like to thank the Isaac Newton Institute for Mathematical Sciences for its hospitality during the programme Periodic and Ergodic Spectral Problems supported by EPSRC Grant Number EP/K032208/1. Both authors thank David Arrowsmith, Matthew Evans and Vito Latora for comments on various versions of this paper and to J.-P. Bouchaud for his comments on the number of stable equilibria in non-gradient systems.
\end{acknowledgments}

%----------------------------------------------------------------------------------------
%	BIBLIOGRAPHY
%----------------------------------------------------------------------------------------

%% PNAS does not support submission of supporting .tex files such as BibTeX.
%% Instead all references must be included in the article .tex document.
%% If you currently use BibTeX, your bibliography is formed because the
%% command \verb+\bibliography{}+ brings the <filename>.bbl file into your
%% .tex document. To conform to PNAS requirements, copy the reference listings
%% from your .bbl file and add them to the article .tex file, using the
%% bibliography environment described above.

%%  Contact pnas@nas.edu if you need assistance with your
%%  bibliography.

% Sample bibliography item in PNAS format:
%% \bibitem{in-text reference} comma-separated author names up to 5,
%% for more than 5 authors use first author last name et al. (year published)
%% article title  {\it Journal Name} volume #: start page-end page.
%% ie,
% \bibitem{Neuhaus} Neuhaus J-M, Sitcher L, Meins F, Jr, Boller T (1991)
% A short C-terminal sequence is necessary and sufficient for the
% targeting of chitinases to the plant vacuole.
% {\it Proc Natl Acad Sci USA} 88:10362-10366.

%% Enter the largest bibliography number in the facing curly brackets
%% following \begin{thebibliography}

\appendix[Supporting Information]

In this appendix we express the mean density of real eigenvalues in the real elliptic ensemble, see  Equation (\ref{EE}) below, of $N\times N$ matrices $X_N$ in terms of the modulus of the spectral determinant of real elliptic matrices of size $(N-1)\times (N-1)$. Our derivation is based on the method suggested in [44], however our Jacobian computation differs from the one given in [44] and may be of interest on its own.   For a similar calculation in the context of complex matrices see [45].

\subsection*{Housholder reflections and partial triangularization of real matrices}

The key idea of  [44] is based on employing Householder reflections described by matrices
\begin{equation}\label{Haush}
P_{\mathbf{v}} = I_N- 2 \mathbf{v} \otimes  \mathbf{v}^T, \quad  \mathbf{v}^T \mathbf{v}=1.
\end{equation}
where $I_N$ is $N\times N$ identity matrix, $ \mathbf{v}$ is column vector in $\mathbb{R}^N$ of unit length. Its transpose,  $\mathbf{v}^T$, is row vector, so that the Kronecker product  $\mathbf{v} \otimes  \mathbf{v}^T$ is a matrix. $P_{\mathbf{v}}$ describes the reflection about the hyperplane with normal $\mathbf{v}$ and passing through the origin. Obviously, $P_{\mathbf{v}}$ is is symmetric and orthogonal: $P_{\mathbf{v}}^T=P_{\mathbf{v}}$ and $P_{\mathbf{v}}^2=I_N$.

Let $\mathbf{e}_1$ be the first vector of the standard basis in $\mathbb{R}^N$, i.e., $\mathbf{e}_1^T=(1,0, \ldots, 0)$. For any unit vector $\mathbf{x}, \, |\textbf{x}|=1$ define
\begin{equation}\label{v}
\mathbf{v}=\frac{\mathbf{x}+\mathbf{e}_1}{\sqrt{2(1+x_1)}}.
\end{equation}
and consider the Hausholder reflection $P_{\mathbf{v}}$ built from the above $\mathbf{v}$ according to (\ref{Haush}).
Then it is easy to check that $P_{\mathbf{v}}\mathbf{x}=-\mathbf{e}_1$. This implies that for any nonzero vector $\mathbf{x}$ there exists a Housholder reflection such that $P_{\mathbf{v}}\mathbf{x}=k\mathbf{e}_1$, with $k=-|\mathbf{x}|$.

Let $\lambda$ be a real eigenvalue of the $N\times N$ matrix $A^{(N)}$ with real entries $A^{(N)}_{ij}$, i.e. $A^{(N)}\mathbf{x}=\lambda \mathbf{x}$ for some column $N-$vector $\mathbf{x}$ of unit length. Our goal is to demonstrate that it is always possible to  represent that matrix as
\begin{equation}\label{reduct}
A^{(N)}=P
\begin{pmatrix}
  \lambda  & {\bf w}^T \\
  0 & A^{(N-1)}
\end{pmatrix} P^T
\end{equation}
for some real $(N-1)$- component vector $\mathbf{w}$ and a real matrix $ A^{(N-1)}$ of size  $(N-1)\times (N-1)$.
To verify this take the Housholder reflection acting on the eigenvector $\mathbf{x}$ as $P_{\mathbf{v}}\mathbf{x}=k\mathbf{e}_1$. Obviously $P_{\mathbf{v}}A^{(N)}\mathbf{x}=\lambda P_{\mathbf{v}}\mathbf{x}=\lambda k \mathbf{e}_1$. In view of $P_{\mathbf{v}}^2=I_N$ we can rewrite the left-hand side as $P_{\mathbf{v}}A^{(N)}P_{\mathbf{v}}\,P_{\mathbf{v}} \mathbf{x}=k\,P_{\mathbf{v}}A^{(N)}P_{\mathbf{v}} \mathbf{e}_1$ which after denoting $\tilde{A}=\,P_{\mathbf{v}}A^{(N)}P_{\mathbf{v}}$ imples $ \tilde{A}\mathbf{e}_1=\lambda \mathbf{e}_1$. Using the definition of $\mathbf{e}_1$ we see that $ \tilde{A}_{11}=\lambda$ and $ \tilde{A}_{1j}=0$ for all $j=2,\ldots,N$. Therefore $\tilde{A}$ can be written as $\tilde{A}=\begin{pmatrix}  \lambda  & {\bf w} \\  0 & A^{(N-1)}\end{pmatrix}$ and as $A^{(N)}=\,P_{\mathbf{v}}\tilde{A}P_{\mathbf{v}}$ the relation (\ref{reduct}) follows.

Considering now the volume element $dA^{(N)}=\prod_{i,j}^{N}dA^{(N)}_{ij}$ our next goal to
write it down in terms of $N^2$ independent variables parametrizing the right-hand side of (\ref{reduct}),
 that is $(N-1)^2$ variables parametrizing $ A^{(N-1)}$, $N-1$ components of $\mathbf{w}$, $1$ parameter for $\lambda$, and the remaining $N-1$ parameters for representing the matrix $P$. A convenient parametrization for $P$ comes from
 employing (\ref{v}) for the vector $\mathbf{v}$, which shows that the last $N-1$ components of that vector
$(v_2,\ldots,v_N)^T\equiv \mathbf{q}$ can be used as independent variables, whereas normalization fixes the first component. Writing $\mathbf{v}^T=\left(\sqrt{1-\mathbf{q}^T\mathbf{q}},\mathbf{q}\right)$ with $|\mathbf{q}|<1$ and employing (\ref{Haush}) yields an explicit parametrization :
\begin{equation}\label{P}
P=\begin{pmatrix}
  2\mathbf{q}^T\mathbf{q}-1  & -2\mathbf{q}^T\sqrt{1-\mathbf{q}^T\mathbf{q}} \\
  -2\mathbf{q}\sqrt{1-\mathbf{q}^T\mathbf{q}} & I_{N-1}-2\mathbf{q}\bigotimes\mathbf{q}^T
\end{pmatrix}
\end{equation}
The problem therefore amounts to calculating the Jacobian of the transformation $A^{(N)}\to (\lambda,\mathbf{w},\mathbf{q},A^{(N-1)})$. To that end we start with differentiating (\ref{reduct}) which gives
\[
dA^{(N)}\!=\!P\left\{\left[(PdP),\begin{pmatrix}
  \lambda  & {\bf w}^T \\
  0 & A^{(N-1)}
\end{pmatrix}\right]\!+\!\begin{pmatrix}
  d\lambda  & d{\bf w}^T \\
  0 & dA^{(N-1)}
\end{pmatrix}\right\}P^T \, ,
\]
where we employed that $dP\,P=-P\,dP$ and used the notation $[A,B]=AB-BA$ for the matrix commutator.
A direct calculation using (\ref{P}) shows that the matrix   $(PdP)$ can be symbolically written as
\begin{equation}\label{PdP}
(PdP)=\begin{pmatrix}
  0  & -d{\bf b}^T \\
  d{\bf b} & dF
\end{pmatrix},
\end{equation}
where the expression for $dF$ is immaterial for our goals, and
\begin{eqnarray}\label{db}
d{\bf b}&=&\left(-2\sqrt{1-\mathbf{q}^T\mathbf{q}}+\frac{1-2\mathbf{q}^T\mathbf{q}}{\sqrt{1-\mathbf{q}^T\mathbf{q}}}
\mathbf{q}\bigotimes\mathbf{q}^T\right)\,d\mathbf{q}
\\ \nonumber
& & +  \left(-4\sqrt{1-\mathbf{q}^T\mathbf{q}}
+\frac{1-2\mathbf{q}^T\mathbf{q}}{\sqrt{1-\mathbf{q}^T\mathbf{q}}}
\right)\left(d\mathbf{q}^T\mathbf{q}\right)\mathbf{q} \, .
\end{eqnarray}
Substituting (\ref{PdP}) into the expression for $dA^{(N)}$ we find that
\[
dA^{(N)}\!\!=\!\!
 P\!\begin{pmatrix}
  d\lambda-\mathbf{w}^Td \mathbf{b}  & d\mathbf{w}^T\!+\!d{\bf b}^T(\lambda-A^{(N-1)})- \mathbf{w}^T\,dF \\
 (\lambda-A^{(N-1)})\, d{\bf b} & dA^{(N-1)}\!+\!d\mathbf{b}\bigotimes \mathbf{w}^T\!+\! [dF,A^{(N-1)}]
\end{pmatrix}\!P^T \, .
\]
From this expression we easily read off the required Jacobian to be given by
\begin{equation}\label{Jac}
\mathbf{Jacobian}=|\det{(\lambda I_{N-1}-A^{(N-1)})}|\,\left|\frac{\partial \mathbf{b}}{\partial\mathbf{q}}\right| \, ,
\end{equation}
where the last factor symbolically denotes the part of the Jacobian coming from
 the transformation $d\mathbf{b}\to d\mathbf{q}$ described in (\ref{db}). A straightforward calculation shows that
\[
\left|\frac{\partial \mathbf{b}}{\partial\mathbf{q}}\right|=2^N(1-\mathbf{q}^T\mathbf{q})^{\frac{N}{2}-1} \, ,
\]
so that finally we arrive at the change-of-measure formula
\begin{eqnarray}\label{volchange}
dA^{(N)}=& & \\ \nonumber
& & \hspace*{-10ex} 2^N(1\!\!-\!\!\mathbf{q}^T\mathbf{q})^{\frac{N}{2}-1}\,|\det \! {(\lambda I_{N-1} \!\!-\!\! A^{(N-1)}\!)}|\,d\lambda \, dw^{N-1}dq^{N-1}dA^{(N-1)} \, .
\end{eqnarray}

\subsection*{Elliptic Ensemble of Gaussian Random Matrices.} The Joint Probability Density (JPD) of the  Elliptic Ensemble of Gaussian random matrices $X_N$ of size $N\times N$ whose entries have covariance
$\left\langle X_{ij}X_{nm} \right\rangle=\delta_{in}\delta_{jm}+  \tau \delta_{jn}\delta_{im}$
is given by
\begin{equation}\label{EE}
{\cal P}(X_N)={\cal Z_N}^{-1}\exp{-\frac{1}{2(1-\tau^2)}\left(\Tr(X_NX_N^T)-\tau\, \Tr(X_N^2)\right)} \, ,
\end{equation}
 where ${\cal Z}_N$ is the corresponding normalization factor
\[
{\cal Z}_N=2^{N/2}\pi^{N(N+1)/2}(1+\tau)^{N(N+1)/4}(1-\tau)^{N(N-1)/4} \, .
\]
Our goal is to find the JPD of variables $\lambda,\mathbf{w},\mathbf{q},X_{N-1}$
used to perform the partial triangulation of $X_N$ via (\ref{reduct}), with the role of $A^{(N)}$ played now by $X_N$. We have
\[
\Tr X_NX_N^T = \lambda^2+\mathbf{w}^T\mathbf{w}+\Tr X_{N-1}X_{N-1}^T, \, \Tr X_N^2 =\lambda^2+\Tr X_{N-1}^2\, .
\]
 Taking into account the change-of-measure formula (\ref{volchange}) we see that the corresponding JPD can be written as
 %\begin{equation}\label{JPD}
\begin{eqnarray*}
{\cal \tilde{P}}(\lambda,\mathbf{w},\mathbf{q},X_{N-1})\!\!&\!\!\!\!=\!\!\!\!&\!\! 2^N(1-\mathbf{q}^T\mathbf{q})^{\frac{N}{2}-1}
e^{-\frac{1}{(1-\tau^2)}\mathbf{w}^T\mathbf{w}} \times \\
& & \hspace*{-10ex}   \frac{{\cal Z}_N}{{\cal Z}_{N-1}} e^{-\frac{\lambda^2}{2(1+\tau)}} \,|\det{(\lambda I_{N-1}-X_{N-1})}| \, {\cal P}(X_{N-1})\, .
\end{eqnarray*}
By definition, the density of real eigenvalues $\rho^{(r)}_N(\lambda)$ is obtained by integrating the above JPD
over variables $|\mathbf{q}|<1,-\infty<\mathbf{w}<\infty$ and finally over $X_{N-1}$. After performing the integrals over $\mathbf{q}$ and $\mathbf{w}$ we immediately arrive at the relation
\[
\rho^{(r)}_N(\lambda)= \frac{1}{{\cal C}_{N-1}(\tau)}\,  e^{-\frac{\lambda^2}{2(1+\tau)}} \left\langle|\det{(\lambda I_{N-1}-X_{N-1})}|\right\rangle_{X_{N-1}}\, ,
\]
where ${\cal C}_{N-1}(\tau)$ is a certain constant which can be found explicitly. This is precisely equivalent to Equation (13) in the main text with the obvious change of notations: $\lambda \to x$ and $N\to N+1$.

%----------------------------------------------------------------------------------------

\end{article}

\end{document}